\documentclass[11pt,a4paper]{article}
\pdfoutput=1

\usepackage{jheppub}
\usepackage{latexsym}
\usepackage{revsymb}
\usepackage{multirow}
\usepackage{color}
\usepackage[usenames,dvipsnames,svgnames,table]{xcolor}

\usepackage{graphicx}
\usepackage{epsfig}  
\usepackage{epsf}    
\usepackage{dcolumn}
\usepackage{bm}
\usepackage{dcolumn}
\usepackage{textcomp}
\usepackage{float}
\usepackage{subfig}
\usepackage{lineno}
\usepackage{hypcap}
\usepackage[]{hyperref}

\hypersetup{
  bookmarks=true,         
  unicode=false,          
  pdftoolbar=true,        
  pdfmenubar=true,        
  pdffitwindow=true,     
  pdfstartview={FitH},    
  pdfsubject={Neutrino Oscillations Phenomenology},   
  pdfnewwindow=true,      
  pdfcreator={RevTeX},
  colorlinks=true,       
  linkcolor=red,          
  citecolor=blue,        
  filecolor=black,      
  urlcolor=blue,           
}

\def\nue{{\nu_e}}
\def\anue{{\bar{\nu}_e}}
\def\numu{{\nu_{\mu}}}
\def\anumu{{\bar{\nu}_{\mu}}}

\newcommand{\eg}{{\it e.g.}}
\newcommand{\ie}{{\it i.e.}}
\newcommand{\etc}{{\it etc.}}

\newcommand{\beq}{\begin{equation}}
\newcommand{\eeq}{\end{equation}}
\newcommand{\beqa}{\begin{eqnarray}}
\newcommand{\eeqa}{\end{eqnarray}}
\newcommand{\tij}{{\theta_{ij}}}
\newcommand{\tx}{{\theta_{12}}}
\newcommand{\ty}{{\theta_{13}}}
\newcommand{\tz}{{\theta_{23}}}
\newcommand{\ta}{{\theta_{14}}}
\newcommand{\tb}{{\theta_{24}}}
\newcommand{\tc}{{\theta_{34}}}

\newcommand{\dcp}{\delta_{\mathrm{CP}}}

\newcommand{\pab}{P(\nu_{\alpha} \rightarrow \nu_{\beta})}
\newcommand{\pabbar}{P(\bar{\nu}_{\alpha} \rightarrow \bar{\nu}_{\beta})}

\newcommand{\pme}{P_{\mu e}}

\newcommand{\pmebar}{P_{\bar{\mu} \bar{e}}}

\newcommand{\da}{\delta_{13}}
\newcommand{\db}{\delta_{24}}
\newcommand{\dc}{\delta_{34}}
\newcommand{\tmet}{\theta^{3\nu}_{\mu e}}
\newcommand{\tmef}{\theta^{4\nu}_{\mu e}}

\preprint{\\ FERMILAB-PUB-15-323-T \\ NSF-KITP-15-102 \\ MITP/15-058}

\title{The impact of  sterile neutrinos on CP measurements at long baselines}

\author[a,c]{Raj Gandhi,}
\author[b]{Boris Kayser,} 
\author[a]{Mehedi Masud,} 
{\author[a]{Suprabh Prakash$\,$}

\affiliation[a]{Harish-Chandra Research Institute, Chhatnag Road, Jhunsi,
Allahabad 211019, India}
\affiliation[b]{Theoretical Physics Department, Fermilab, P.O. Box 500, Batavia,
IL 60510 USA}
\affiliation[c]{Neutrino Division, Fermilab, P.O. Box 500, Batavia,
IL 60510 USA}

\emailAdd{raj@hri.res.in}
\emailAdd{boris@fnal.gov}
\emailAdd{masud@hri.res.in}
\emailAdd{suprabhprakash@hri.res.in}

\begin{abstract}
{With the Deep Underground Neutrino Experiment (DUNE) as an example, we show that 
the presence of even one sterile neutrino of mass $\sim$1 eV can significantly impact 
the measurements of CP violation in long baseline experiments. Using a probability 
level analysis and neutrino-antineutrino asymmetry calculations, we discuss the large magnitude of these  
effects, and show how they translate into significant event rate deviations at DUNE. 
Our results demonstrate that measurements which, when interpreted in the context of 
the standard three family paradigm, indicate CP conservation at long baselines, may, 
in fact hide large CP violation if there is a sterile state. Similarly, 
any data indicating the violation of CP cannot be properly interpreted within the 
standard paradigm unless the presence of sterile states of mass O(1 eV) can be 
conclusively ruled out. Our work underscores the need for a parallel and linked 
short baseline oscillation program and a highly capable near detector for DUNE,
in order that its highly anticipated results on CP violation in the lepton sector 
may be correctly interpreted.
}
\end{abstract}

\keywords{}
\arxivnumber{}

\begin{document}
\maketitle
\flushbottom

\section{Introduction}
\label{sec:intro}
A major goal of present and future long-baseline neutrino oscillation
experiments is to establish that leptons violate CP, or else to place a
stringent upper limit on any such violation.  (For recent status reviews see
\cite{thomson_winp2015,1.4915573,1.4915574,1.4915575,1.4915576,1.4915579,
1.4915580,1.4915581,1.4915582,1.4915583,1.4915584,deGouvea:2013onf}.) Our thinking about
these experiments usually assumes the standard neutrino paradigm, in which there
are  just three neutrino mass eigenstates separated by just two independent
mass-squared splittings, three mixing angles $\tij$, and just
one CP-violating phase $\dcp$ relevant to oscillation. However, a variety of
short-baseline anomalies \cite{Aguilar:2001ty,
AguilarArevalo:2008rc, Mention:2011rk, Mueller:2011nm, Aguilar-Arevalo:2013pmq}
hint at the possible existence of short-wavelength oscillations, driven by one
or more $O$(1 eV$^2$) mass-squared splittings that are much larger than the two
splittings of the standard paradigm. These short-wavelength oscillations are
purportedly already  significant when the (Travel distance $L$)/(Energy $E$) of
neutrinos in a beam is only  $\sim$ 1 km/GeV. Of course, they are still present at the far
detector of any long-baseline experiment, where $L/E$ is, say, $\sim$ 500 km/GeV. In
this work, we have explored the consequences of the short-wavelength
oscillations - should they be real - for measurements at long baselines,
especially measurements that probe CP violation. We find that these consequences
could be considerable. For example, it is possible for long-baseline results,
interpreted without taking the short-wavelength oscillations into account, to
imply that CP violation is very small or totally absent, when in reality it is
very large. In 
addition, long-baseline measurements interpreted as determining the sole
oscillation-relevant CP-violating phase in the standard paradigm could in fact
be measuring something else. 

The large splittings hinted at by the short-baseline anomalies imply the
existence of additional, largely sterile, neutrino mass eigenstates, beyond the
three of the standard scenario (referred to as 3+0 in what follows). These
additional mass eigenstates introduce not only additional splittings but also
additional mixing angles and phases. For simplicity, we restrict ourselves to
the scenario, referred to as 3+1, with only one additional mass eigenstate. In
this scenario, there are six mixing angles, and three CP-violating phases that
can affect oscillation. Denoting the mass eigenstates of 3+0, as usual, as
$\nu_1, \nu_2, \nu_3$, and the additional mass eigenstate as $\nu_4$, and
defining the mass-squared splittings as $\delta m^{2}_{ij} = m^{2}_{i} -
m^{2}_{j}$ (i, j = 1, 2, 3, 4 \& $i
\neq j$), we have, according to present  data,
\beq \label{eq:sterile_large_mass}
\delta m^2_{41} \sim \delta m^2_{42} \sim \delta m^2_{43} >> |\delta m^2_{31}|\sim|\delta m^2_{32}|
>> \delta m^2_{21}.
\eeq

Since the probability of an oscillation driven by a splitting $\delta m_{ij}^{2}$  is proportional
to $\sin^2\Delta_{ij}$, where $\Delta_{ij} = 1.27 \times 
\frac{\delta m^{2}_{ij}[\textrm{eV}^{2}] \times L [km]}{E [GeV]}$ , when L/E $\sim$
500 km/GeV, the short-wavelength oscillations driven by the large splittings
involving $\nu_4$ will be averaged to an L/E-independent value by the finite
energy resolution of any realistic detector. But these rapid oscillations are
still present and can have a major impact. 

We perform our calculations for the 3+1 scenario as manifested in the proposed
Deep Underground Neutrino Experiment (DUNE)\footnote{
The inputs we use, and the corresponding references, pertain  to the
erstwhile Long Baseline Neutrino Experiment (LBNE), which  has  undergone  a new
phase of
internationalisation and expansion. This has  led to a change in the name of the
experiment, to DUNE.  Nonetheless, it is expected that the configuration we
assume here vis a
vis fluxes, baseline and energies will remain largely
intact\cite{Hewett:2012ns,Adams:2013qkq,Bass:2013vcg}.}. While we do not explore
3+N scenarios with N$>$1, we expect that if the consequences of having one extra
neutrino for long-baseline measurements are substantial, those of having more
than one must be substantial as well, since the world with one extra neutrino is
in a sense a special case of that with more than one. 

Previous work examining the effects of sterile neutrinos at long baselines
includes several studies of neutrino factories feeding baselines of about
3000 km - 7500 km, with muon energies in the range 20 GeV - 50 GeV, focussing on
effects at both near and far detectors \cite{Donini:2001xp, Dighe:2007uf,
Donini:2008wz, Yasuda:2010rj, Meloni:2010zr}. More recent work
\cite{Klop:2014ima} includes a study of effects relevant to T2K
\cite{Abe:2013hdq} and a combined study \cite{Bhattacharya:2011ee} for T2K,
MINOS \cite{Adamson:2014vgd} and reactor experiments. Additionally, issues
having some overlap with those addressed here  for DUNE have been discussed in
\cite{Hollander:2014iha}, and, very recently, in \cite{Berryman:2015nua}.

In Sec.\ \ref{sec:probability} we examine, in the 3+1 scenario, the probability $P_{\mu e}^{4\nu}$
of $\nu_e$  appearance at the far end of a long-baseline beam of neutrinos born
as $\nu_{\mu}$  . We derive an analytical expression for  $P_{\mu e}^{4\nu}$ 
valid in vacuum, and discuss how its qualitative features change when matter
effects, which will be large in the case of DUNE, are taken into account.
Sec.\ \ref{sec:acp} focuses on the neutrino-antineutrino asymmetries that are possible at
long baselines in the 3+0 and 3+1 scenarios, and on how the possible asymmetries
in these two scenarios compare. Sec.\ \ref{sec:event_rate} presents long-baseline far detector
event rates, obtained by performing realistic rate calculations for the two
scenarios. Sec.\ \ref{sec:implic} briefly discusses two important implications of the results
presented here. In Sec.\ \ref{sec:summary}, we summarize and  discuss what conclusions could
possibly  be drawn, and what ones could not be drawn, especially concerning CP
violation, from a given set of  long-baseline oscillation results, and conclude.

\section{The $3+1$ electron appearance probability in vacuum and matter}
\label{sec:probability}

For CPV discovery in long baseline super-beam experiments, the electron neutrino
appearance probability $\pme$ is crucial. We discuss its analytic form in vacuum
for  the $3+1$ scenario prior to discussing  the matter case. While it is the
latter that is relevant for DUNE in particular, and other long baseline (LBL) experiments at
baselines of $O(1000)$ km in general, the form of the vacuum expression provides
a useful template for the identification of terms the importance of which will
be accentuated by the presence of matter.

We adopt the following parameterisation\footnote{This choice, which, at first, appears not to be the most general one which could be made,  is motivated by the fact that in any parameterisation, the electron neutrino 3+1  appearance probability in vacuum  turns out to be dependent on only  two specific linear combinations of the three independent phases. For simplicity, we have thus incorporated this at the outset using the fact that 
the first and second row elements ($U_{ei}$
and $U_{\mu i}$) in $\text{U}_{\text{PMNS}}^{3+1}$ will not have $\theta_{34}$ or
$\delta_{34}$ in
them.} for the PMNS matrix
in the presence of a sterile neutrino,

\beq
U_{\text{PMNS}}^{3+1}=O(\theta_{34},
\delta_{34})O(\theta_{24},\delta_{24})O(\theta_{14})O(\theta_{23})O(\theta_{13},
\delta_{13})O(\theta_{12}).
\eeq
Here,  in general, $O(\theta_{ij},\delta_{ij})$ is a rotation matrix
in the $ij$ sector with associated phase $\delta_{ij}$. For example,
\begin{equation*}
O(\theta_{24},\delta_{24}) = 
\begin{pmatrix}
1 & 0 & 0 & 0 \\
0 & \cos\tb & 0 & e^{-i\delta_{24}}\sin\tb \\
0 & 0 & 1 & 0 \\
0 & -e^{i\delta_{24}}\sin\tb & 0 & \cos\tb
\end{pmatrix};
O(\theta_{14}) = 
\begin{pmatrix}
\cos\ta & 0 & 0 & \sin\ta \\
0 & 1 & 0 & 0 \\
0 & 0 & 1 & 0 \\
-\sin\ta & 0 & 0 & \cos\ta
\end{pmatrix} \text{\etc}
\end{equation*}
Using the standard formula for a flavour transition oscillation probability, we
have, for the 3+1 case:
\beqa \label{eq:pme_gen}
\nonumber P_{\mu e}^{4\nu} &=& 4|U_{\mu4}U_{e4}|^2\times0.5 \\
\nonumber &-& 4Re(U_{\mu1}U_{e1}^*U_{\mu2}^*U_{e2})\sin^2\Delta_{21} +
2Im(U_{\mu1}U_{e1}^*U_{\mu2}^*U_{e2})\sin2\Delta_{21}\\
\nonumber &-& 4Re(U_{\mu1}U_{e1}^*U_{\mu3}^*U_{e3})\sin^2\Delta_{31} +
2Im(U_{\mu1}U_{e1}^*U_{\mu3}^*U_{e3})\sin2\Delta_{31}\\
 &-& 4Re(U_{\mu2}U_{e2}^*U_{\mu3}^*U_{e3})\sin^2\Delta_{32} +
2Im(U_{\mu2}U_{e2}^*U_{\mu3}^*U_{e3})\sin2\Delta_{32}.
\eeqa

In arriving at the above expression, we have only  assumed (based on Eq.\ \ref{eq:sterile_large_mass}  above) that $\sin^2\Delta_{4i}$ averages out to be 0.5 at long
baselines, and similarly
$\sin2\Delta_{4i}$ averages out to be 0, when $i=1,2,3$.

After substituting the values of the $U_{\alpha i}$ in terms of the mixing
angles, we obtain :
\beqa \label{eq:p_me}
\nonumber P_{\mu e}^{4\nu} &=& \frac{1}{2}\sin^22\tmef \\
\nonumber &+&  (a^2\sin^22\tmet - \frac{1}{4}\sin^22\ty\sin^22\tmef)
\big[\cos^2\tx\sin^2\Delta_{31}+\sin^2\tx\sin^2\Delta_{32} \big]\\
\nonumber &+& \cos(\da) ba^2  \sin2\tmet 
\big[\cos2\tx\sin^2\Delta_{21}+\sin^2\Delta_{31}-\sin^2\Delta_{32} \big]\\
\nonumber &+& \cos(\db)  ba \sin2\tmef 
\big[\cos2\tx\cos^2\ty\sin^2\Delta_{21}-\sin^2\ty(\sin^2\Delta_{31}
-\sin^2\Delta_{32}) \big]\\
\nonumber &+& \cos(\da + \db)  a \sin2\tmet \sin2\tmef
\big[-\frac{1}{2}\sin^22\tx\cos^2\ty\sin^2\Delta_{21} \\ 
\nonumber &+& \cos2\ty(\cos^2\tx\sin^2\Delta_{31}+\sin^2\tx\sin^2\Delta_{32})
\big]\\
\nonumber &-& \frac{1}{2} \sin(\da) ba^2 \sin2\tmet 
\big[\sin2\Delta_{21}-\sin2\Delta_{31}+\sin2\Delta_{32} \big]\\
\nonumber &+& \frac{1}{2} \sin(\db) ba \sin2\tmef 
\big[\cos^2\ty\sin2\Delta_{21}+\sin^2\ty(\sin2\Delta_{31}-\sin2\Delta_{32})
\big]\\
\nonumber &+& \frac{1}{2} \sin(\da + \db) a \sin2\tmet \sin2\tmef 
\big[\cos^2\tx\sin2\Delta_{31}+\sin^2\tx\sin2\Delta_{32} \big]\\
\nonumber &+&
(b^2a^2-\frac{1}{4}a^2\sin^22\tx\sin^22\tmet-\frac{1}{4}
\cos^4\ty\sin^22\tx\sin^22\tmef) \sin^2\Delta_{21},\\
\eeqa
where,
\beqa 
\sin2\theta_{\mu e}^{3\nu}=\sin2\ty\sin\tz\\
b=\cos\theta_{13}\cos\theta_{23}\sin2\tx\\
\sin2\theta_{\mu e}^{4\nu}=\sin2\theta_{14}\sin\theta_{24}\\
a=\cos\theta_{14}\cos\theta_{24}.
\eeqa 

Prior to proceeding, we briefly discuss the allowed ranges for the $3+1$ mixing
angles that we have used in our calculations. These have been obtained using the
results of \cite{Kopp:2013vaa}, which  takes all available data on
short-baseline (SBL) oscillations and performs a global fit to constrain
active-sterile mixing.
$|U_{e4}|^{2}$ is constrained by $\nue$ and $\anue$ disappearance searches, and
is  equal to $\sin^{2}\ta$. 
 From  \cite{Kopp:2013vaa}, the 99\% C.L. limit can be
taken to be (with some extrapolation, as the result is given for 95\%),
$$|U_{e4}|^{2}  \in  [0,0.1],$$
which gives
$$\ta  \in  [0, 20^{o}].$$
Similarly, using $\numu, \anumu$ and neutral current disappearance searches, one
can constrain $U_{\mu 4}$ and $U_{\tau 4}$, which are given by,
$$|U_{\mu 4}| = \cos\ta \sin\tb,
|U_{\tau 4}| = \cos\ta \cos\tb \sin\tc.$$ The constraints on these elements are found to be,
$$|U_{\mu 4}|^{2} \in [0, 0.03], 
|U_{\tau 4}|^{2} \in [0, 0.3].$$ These translate to, 
$$\tb \in [0, 11^{o}], \tc \in [0, 31^{o}].$$

Apropos Eq.\ \ref{eq:p_me}, we note that, as expected,  
the vacuum appearance  probability is independent of the 3-4 mixing angle and
the
associated CP phase. This important simplification, however, does not carry over to the matter
case, as we show below.
Secondly, Eq.\ \ref{eq:p_me} contains terms proportional to the sines and cosines of a) the
$3+1$ CP phase $\delta_{24}$, and b) the sum $(\delta_{13} + \delta_{24})$. These are interference terms, involving mixing angles from both the  3+0 and the 3+1 sector. In particular, as can be determined by  inspection, the terms involving the sine and cosine of the sum of $\delta_{13}$ and $\delta_{24}$ can be significantly large and lead to appreciable changes in both the amplitude of the overall probability and the extent of CP violation.
These  contributions  become all the more significant once matter effects are
large. We discuss this in more detail below,  both in this section and the next.

\begin{figure}[H]
\centering
\includegraphics[width=0.45\textwidth]
{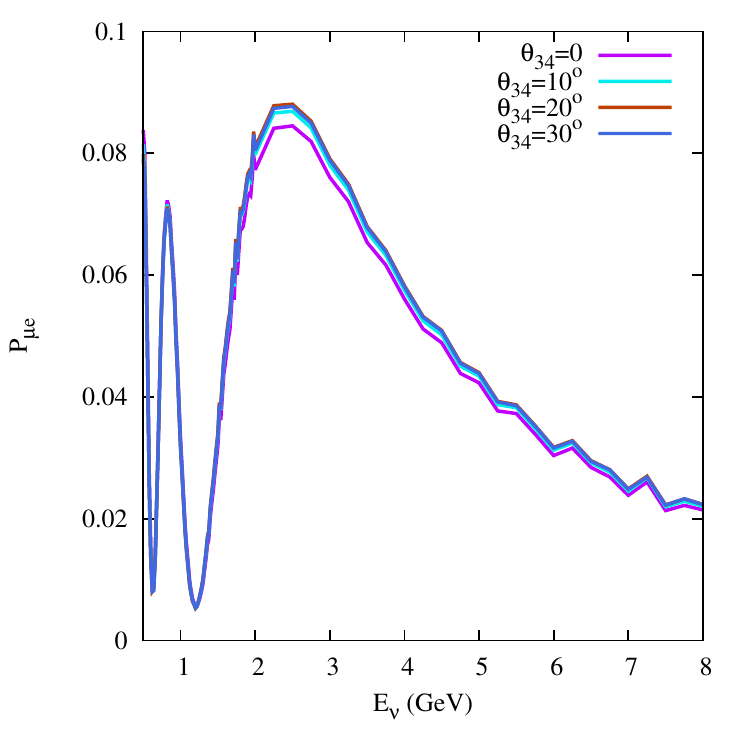}
\includegraphics[width=0.45\textwidth]
{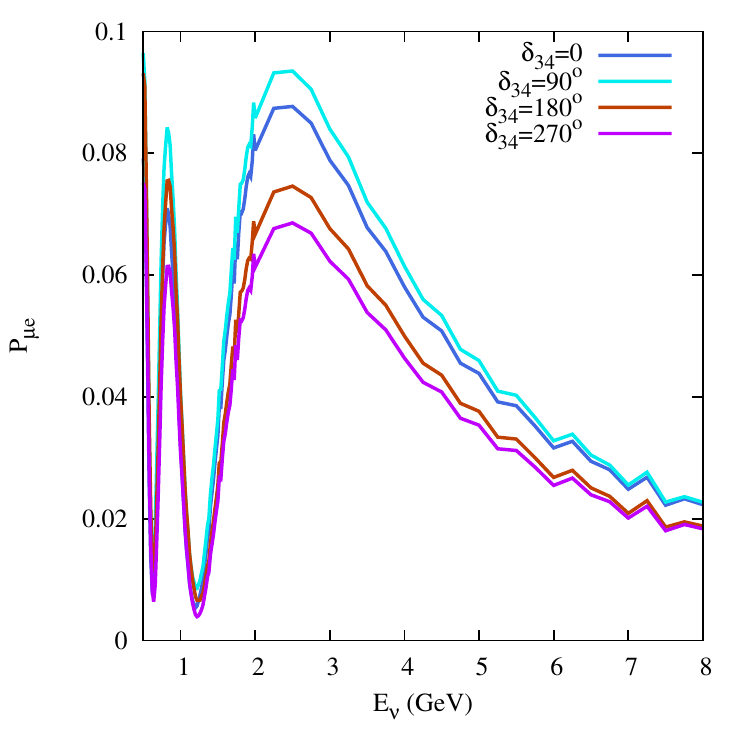}
\caption{\footnotesize{$P_{\mu e}$ vs $E_{\nu}$ in earth matter for
1300 km. Averaging has been done for $\Delta m^2_{4i}$ induced oscillations. 
In the left panel, the effect of varying  $\theta_{34}$ within its allowed range
is shown with all the 
CP phases kept equal to 0.
In the right panel, we show the effect of varying CP violating phase
$\delta_{34}$ when $\theta_{34}=30^\circ$, and the other phases are 0. For both panels, we set 
$\theta_{14} = 20^{\circ}$ and $\theta_{24} = 10^{\circ}$, and the parameters related
to the 3+0 sector at the best-fit values specified in Sec. \ref{sec:event_rate}}.}
\label{fig:3-4prob}
\end{figure}

The matter eigenstates
bring about  a  dependence  on all mixing angles and phases. Specifically,
unlike the vacuum case, the 3-4 mixing angle and its associated phase are no
longer quiescent, and the 3+1 electron neutrino appearance probability exhibits
a significant
dependence on them. This is illustrated by Fig.\ \ref{fig:3-4prob}, where we have used the
General Long Baseline Experiment Simulator (GLoBES) \cite{Huber:2004ka,
Huber:2007ji} to generate the plots. The left panel shows the variation of 
$P_{\mu e}^{4\nu}$ with energy for no CP violation (all three Dirac CP phases
set to zero) and  four different values of the  3-4
mixing angle.  While the variation due to 
$\theta_{34}$, shown in the left panel, is not very large, the effect of
varying 
$\delta_{34}$ (while keeping $\theta_{34}$ fixed, right panel) within its
allowed range is quite significant. It is also striking that these large effects
on the probability  in the presence of matter are brought about by parameters
which are completely absent in the vacuum expression (Eq.\ \ref{eq:p_me})
and which (in our chosen parametrization) play no role in SBL situations. They
illustrate the important role matter plays in invoking and  enhancing the
effects of sterile states  at long baselines. In these plots,  all other
parameters are fixed as described in the caption. 

\begin{figure}[H]
\centering
\includegraphics[width=0.95\textwidth]
{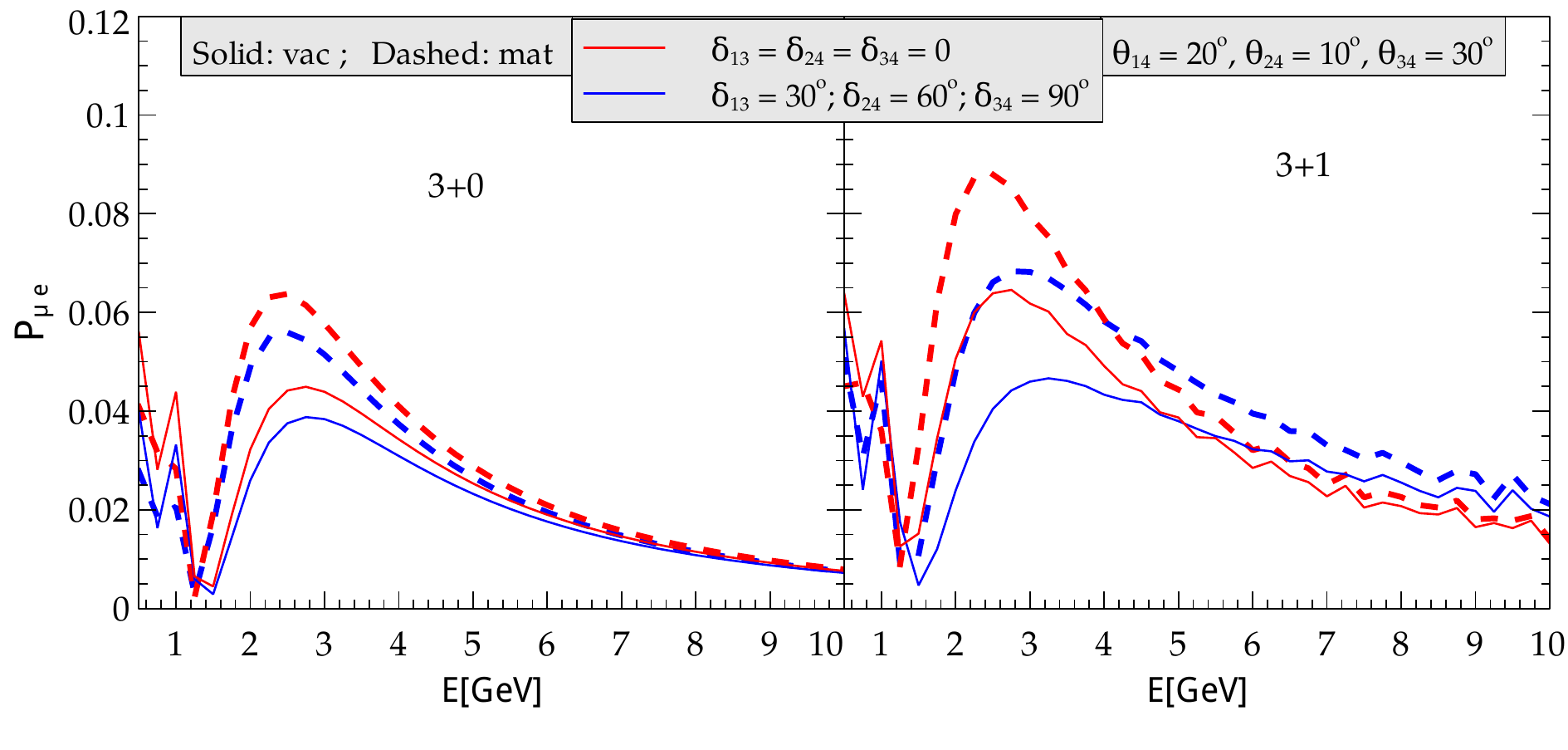}
\caption{\footnotesize{$\pme$ (both for vacuum and matter) for 3+0 (left
panel) and 3+1 (right panel) vs.\ energy.
 The red curves represent the CP conserving case, while the blue ones depict the
case
  with phases set to non-zero fixed values  (see the plot label). For the blue
curve in the left panel, the sole 3+0 phase
  $\dcp$ was taken as $30^{o}$. Normal hierarchy is taken to be the  true
hierarchy here, and parameters related to the 3+0 sector have been set at
the best-fit values specified in Sec.\ \ref{sec:event_rate}}.}
\label{fig:std_sterile_comp}
\end{figure}

Fig.\ \ref{fig:std_sterile_comp} emphasizes the dependencies discussed above from a
slightly more general
perspective. 
In the right (3+1) panel,
the significant differences between the solid and dashed lines
of a given colour emphasize the role played by matter, while the equally
significant differences between the blue  and red
dashed (solid)  lines demonstrate the important role played by 
CP violating phases at long baselines in matter (vacuum) if they are non-zero.
Turning to the left (3+0) panel, we
note the relatively large differences between these curves and their
counterparts in the right panel, underlining the significant effects of sterile
neutrinos at the operational baseline for DUNE.

In summary, Figs \ref{fig:3-4prob} and \ref{fig:std_sterile_comp} demonstrate that  additional CP phases related to a
eV$^2$ sterile sector play an important role at long baselines. Their effects
are heightened  by the presence of matter. In addition, parameters related to
the sterile sector which are dormant at short baselines and in vacuum-like
conditions are no longer inert once baselines are long and matter effects are
important. 
 
\section{A discussion of Neutrino-Antineutrino asymmetries in matter}
\label{sec:acp}

The consideration of CP violation in terms of an asymmetry defined at the
probability level provides additional insight into the conclusions which can be
reliably drawn from data if we do not know whether 3+0 or 3+1 is the choice
nature has made.
Consider the asymmetry defined as,

\begin{equation}\label{eq:A_cp}
A_{\nu{\bar{\nu}}}^{\alpha\beta} = \frac{\pab - \pabbar}{\pab +
\pabbar}\equiv\frac{\Delta
P_{\alpha\beta}}{\pab + \pabbar}.
\end{equation}

We begin by noting an important difference between the 3+0 and 3+1 scenarios
with respect to the numerator $\Delta P_{\alpha \beta}$ of 
$A_{\nu{\bar{\nu}}}^{\alpha\beta}$. In vacuum, CPT invariance implies that 
$P(\nu_{\beta} \rightarrow \nu_{\alpha}) = P(\bar{\nu}_{\alpha} \rightarrow \bar{\nu}_{\beta})$,
which in turn implies that $\Delta P_{\beta \alpha} = -\Delta P_{\alpha \beta}$, 
and in particular that $\Delta P_{\alpha \beta} = 0$ when $\beta = \alpha$. Thus, when there 
are only three neutrino flavors, there are only three independent potentially 
non-zero CP-violating differences $\Delta P_{\alpha \beta}$ to be measured: $\Delta P_{e \mu}$, 
$\Delta P_{\mu \tau}$ and $\Delta P_{\tau e}$. Now, conservation of probability implies that 
for any number of flavors, 
\begin{equation*}
 \sum_{\beta}\pab = 1 \text{\quad and \quad} \sum_{\beta}\pabbar = 1,
\end{equation*}
where the sums are over all $\beta$, including $\beta = \alpha$. It follows that 
$\sum_{\beta}\Delta P_{\alpha \beta} = 0$. Then, since $\Delta P_{\alpha \beta} = 0$ 
when $\beta = \alpha$, we conclude that in vacuum,
\begin{equation}\label{eq:acp_constraint}
\sum_{\beta \neq \alpha}\Delta P_{\alpha \beta} = 0. 
\end{equation}
When there are only three flavors, this constraint implies that $\Delta P_{e \mu} + 
\Delta P_{e \tau} = 0$ and that $\Delta P_{\mu e} + \Delta P_{\mu \tau} = 0$. Since 
$\Delta P_{\beta \alpha} = -\Delta P_{\alpha \beta}$, it follows that,
\begin{equation}\label{eq:acp_equal_channel}
\Delta P_{e \mu} = \Delta P_{\mu \tau} = \Delta P_{\tau e}.
\end{equation}
That is, the three ``independent'' CP-violating differences are equal. In particular, 
if there are only three flavors, it is not possible for CP invariance to hold in one 
oscillation channel, such as $\overset{(-)}{\nu_{\mu}} \rightarrow \overset{(-)}{\nu_{e}}$, 
and yet be violated in another channel, such as 
$\overset{(-)}{\nu_{\mu}} \rightarrow \overset{(-)}{\nu_{\tau}}$.

This situation changes when there are more than three flavors. For \eg, when there are four 
flavors, as in the 3+1 scenario, there are six independent potentially non-zero differences $\Delta P_{\alpha \beta}$: $\Delta P_{e \mu}$, $\Delta P_{\mu \tau}$, $\Delta P_{\tau e}$, 
$\Delta P_{es}$, $\Delta P_{\mu s}$ and $\Delta P_{\tau s}$, where $s$ refers to the sterile 
flavor. Now the constraint of Eq.\ \ref{eq:acp_constraint} gives rise only to relations like
\begin{equation}\label{eq:acp_sterile_equal_channel}
 \Delta P_{e \mu} = \Delta P_{\mu \tau} + \Delta P_{\mu s}.
\end{equation}
It is now perfectly possible for $\Delta P_{\mu e} (= -\Delta P_{e \mu})$, the CP-violating 
difference that will be the first to be probed experimentally, to be zero, while the 
differences $\Delta P_{\mu \tau}$ and $\Delta P_{\mu s}$ in other oscillation channels 
that are challenging to study, are large\footnote{We note that any long baseline experiment involves
earth-matter effects, which break CPT (in addition to CP). Such breaking is
extrinsic, and due to the asymmetry of the earth matter through which the  neutrinos propagate. While this may appear to destroy the conclusions reached above, which depend on CPT
invariance, this is not the case  as long as an experiment seeks to measure
intrinsic (\ie driven by phases in the mixing matrix) CP violation and devises 
appropriate means to do so.}.

\begin{figure}[H]
\centering
\includegraphics[width=0.85\textwidth]
{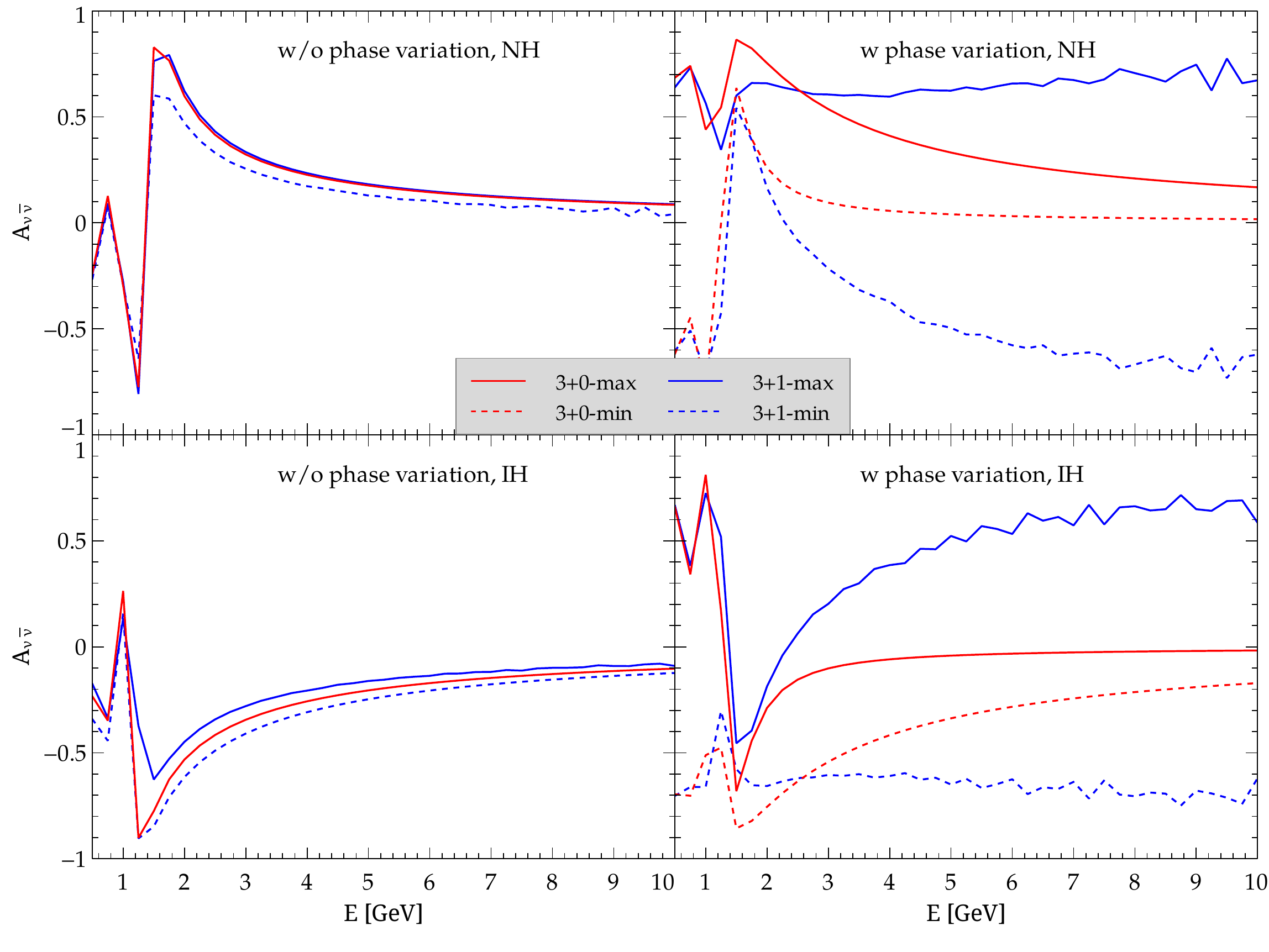}
\caption{\footnotesize{The neutrino-antineutrino asymmetry $A_{\nu \bar{\nu}}$ vs.\ 
energy E. See text for explanation and discussion.}}
\label{fig:ACPplots}
\end{figure}

In Fig.\ \ref{fig:ACPplots}, we show the 
spread of  $A_{\nu{\bar{\nu}}}$\footnote{Henceforth we drop the superscripts
$\alpha$ and $\beta$ and take $A_{\nu{\bar{\nu}}}$ to denote the asymmetry for
$\alpha = \mu$ and  $\beta = e$.} 
at $L = 1300$ km for cases chosen to illustrate some of the important features 
that arise due to the presence of a fourth, sterile state. The left-hand panels 
were created with all CP-violating phases set to zero, so the asymmetries shown in 
these panels are from matter effects only. The right-hand panels were created 
allowing the sole 3+0 CP phase $\dcp$ to vary over its entire physical range in 
the case of 3+0, and the three CP phases $\da$, $\db$ and $\dc$ to vary over their 
entire ranges in the case of 3+1. Thus, these panels show the impact of intrinsic 
CP violation. In all panels, the red curve(s) are for the 3+0 case, and the blue 
ones for the 3+1 case. The top two panels assume a normal hierarchy, and the 
bottom two an inverted hierarchy. In creating all panels, the mass splittings and 
mixing angles of the 3+0 sector were set to the best-fit values specified in Sec.\ 
\ref{sec:event_rate}, the splitting $\Delta m^{2}_{41}$ of the 3+1 sector was set to 
1 $\text{eV}^{2}$, and the 3+1 mixing angles $\ta$, $\tb$ and $\tc$ were varied over their 
allowed ranges. 

With one exception (see below), to create the curves in each panel of Fig.\ \ref{fig:ACPplots} 
for each of the two scenarios, 3+0 and 3+1, we varied the corresponding parameters 
until we found the parameter set that maximizes (minimizes) the energy-integrated 
asymmetry $A_{\nu{\bar{\nu}}}$ for that scenario. The energy-dependent asymmetry 
was then plotted vs.\ energy for this parameter set as a solid (dashed) curve. 
(Note that since it is the energy-integrated asymmetry that is being extremized, 
it is possible for the 3+0 energy-dependent asymmetry to be more extreme than that 
for 3+1 for a limited range of energy, despite the fact that the 3+0 scenario is in 
a sense, a special case of 3+1.) The one exception to our procedure is that, 
since the 3+0 sector parameters other than $\dcp$ were held fixed throughout, 
in creating the left-hand panels, no 3+0 parameters were varied, so there is only a 
single curve, shown as solid, for 3+0.

From the left-hand panels of Fig.\ \ref{fig:ACPplots}, we see that when CP is conserved, the 
neutrino-antineutrino asymmetry vs.\ energy is quite similar in the 3+0 and 3+1 
scenarios. In the 3+1 scenario, this asymmetry is confined to a rather narrow band 
as the 3+1 mixing angles are varied. Although it is not shown, we have found that 
it is confined to a similar narrow band in the 3+0 scenario as the 3+0 parameters 
$\theta_{23}$ and $\Delta m^{2}_{31}$ are varied within their experimental uncertainties. 
Clearly, if an experiment were to measure an asymmetry vs.\ energy that consistently 
lies outside the similar, narrow 3+0 and 3+1 bands that correspond to CP conservation, 
we would have evidence that CP is violated so long as nature has chosen either the 3+0 or 
3+1 scenario. However, a measured asymmetry between $\nu_{\mu} \rightarrow \nu_{e}$ 
and $\bar{\nu}_{\mu} \rightarrow \bar{\nu}_{e}$ that lies within these similar narrow 
bands would not unambiguously signal that there is no CP violation in neutrino oscillation. 
As explained above, when there are more than three flavors, as in the 3+1 scenario, 
it is possible for there to be little or no CP violation in one oscillation channel, 
and yet a large CP violation in some other channel. In addition, for either the 3+0 or 
3+1 case, it might happen that for some non-zero values of the CP-violating phases and 
mixing angles slightly different from those corresponding to the CP-conserving bands, 
the asymmetry still lies within those bands within uncertainties. 

The right-hand panels in Fig.\ \ref{fig:ACPplots} show that when intrinsic CP is violated, 
$A_{\nu \bar{\nu}}$ can be anywhere in a large range. Moreover, for 3+1, this range is much 
larger than for 3+0, and includes almost all of the 3+0 range. Thus, we see that sterile 
neutrinos with $O(1)$ $\text{eV}^{2}$ masses can very substantially impact CP-violation measurements 
at long baselines. While a measured asymmetry outside the band allowed for 3+0 would be 
evidence for new physics beyond 3+0, one inside that band would leave uncertain the precise 
origin of the observed CP violation.

To understand why the effect of a fourth, sterile neutrino on CP violation at long baselines 
can be so large, one notes that CP-violating phases affect physics through interferences 
between amplitudes. As pointed out in \cite{Klop:2014ima}, around the first maximum of the 
atmospheric-wavelength oscillation, where the long-baseline experiments work, or will work, 
the \textit{(new, short wavelength oscillation) - (atmospheric-wavelength oscillation)} interference, 
and the \textit{(atmospheric-wavelength oscillation) - (solar-wavelength oscillation)} interference, 
can easily be of comparable size. Then, if the CP phases are right, 3+1 can be quite different 
from 3+0. 

In the next section, we see how the probability-level results of this section and the 
previous one translate into observable consequences for DUNE by calculating event rates. 

\section{Event Rates at DUNE in the 3+1 and 3+0 scenarios}\label{sec:event_rate}
\label{sec:events}

Having discussed some of the salient features of the probability $P_{\mu
e}^{4\nu}$ both in vacuum and matter at baselines characteristic of DUNE, we now
perform event rate calculations by realistically simulating the experiment.

We recall that DUNE (with specifications very similar to
LBNE,\cite{Hewett:2012ns},
\cite{Adams:2013qkq}) 
is a proposed future super-beam experiment, to be located in the United States,  with a main aim of
establishing or refuting the existence of CPV in the 
leptonic sector. In addition to this primary goal, this facility will also be
able to resolve  other important issues like the mass hierarchy and 
the octant of $\tz$. The ${\nu_{\mu}(\bar{\nu}_{\mu}})$ super-beam 
will originate at the Fermilab. The primary beam simulation 
assumes a 1.2 MW - 120 GeV proton beam which will deliver 
$10^{21}$ protons-on-target (POT) per year. A 35-40 kt Liquid Argon 
(LAr) far-detector will be housed at the Sanford Underground Research Facility 
in the Homestake mine in South Dakota, 1300 km
away. The experiment plans to have a total of 10 years of running, divided
equally 
between neutrinos and anti-neutrinos,  corresponding  to a total
exposure of $35\times10^{22}$ kt-POT-yr. The other experimental details
such as signal and background definitions as well as the detector efficiencies 
taken in this work are the same as those in \cite{Bass:2013vcg}, except with
the  difference that we have not considered tau events in the 
backgrounds. 
The detector efficiencies for both
$\pme$ and $\pmebar$ events are close to 80\% with somewhat less
efficiency for $\pmebar$.

In order to facilitate the drawing of physics conclusions, we assume certain
values and ranges for neutrino oscillation parameters in the standard 3-flavour
paradigm,  which  are  motivated by their current measured ranges and best fit
values. Specifically, 
\begin{itemize}
 \item $\tx$ and $\ty$ have been fixed at $33.48^\circ$ and $8.5^\circ$
respectively \cite{Gonzalez-Garcia:2014bfa}. 
\item We assume that the 2-3 mixing is near-maximal i.e.
$\tz=45^\circ$.\footnote{
We note that the currently-allowed $3\sigma$ range on
$\tz$ is $[38.3^\circ,53.3^\circ]$ with the best fit at 
$42.3^\circ (49.5)^\circ$ for NH (IH) \cite{Gonzalez-Garcia:2014bfa}.
The $\tz$ best fit values from the global analyses \cite{Capozzi:2013csa,
Forero:2014bxa} are different from \cite{Gonzalez-Garcia:2014bfa}.
However, in this work, we make the simplifying assumption of maximal mixing,
relegating a more rigorous statistical analysis to a follow-up work. }
\item We fix $\Delta m^2_{21}$
at $7.5\times10^{-5}~\rm{eV}^2$ and $\Delta m^2_{31}$ at
$2.457\times10^{-3}~\rm{eV}^2$
($-2.374\times10^{-3}~\rm{eV}^2$) for NH (IH). These particular choices  have
been taken from the analysis of global data \cite{Gonzalez-Garcia:2014bfa}. 
\end{itemize}
We note that as stated above, standard global analyses (\cite{Capozzi:2013csa},
\cite{Forero:2014bxa}, \cite{Gonzalez-Garcia:2014bfa}) assume the 3+0 scenario,
but, as has been demonstrated in $\eg$ \cite{Esmaili:2013yea}, their conclusions
 remain very
robust in the presence of sterile neutrinos. Our assumed ranges for the sterile 
sector mixing angles corresponding to the 3+1 scenario are as stated in Sec.\
2. In addition,  
 we assume 
  $\Delta m^2_{41}$ to be $1~\rm{eV}^2$, and vary 
  $\da$, $\db$ and $\dc$ 
        for 3+1 and $\dcp$ for 3+0 in the entire possible range of 
        $[-180^\circ, 180^\circ]$. Finally, we use the fluxes provided in
\cite{mbishai}.

\begin{figure}[H]
\centering
\includegraphics[width=0.45\textwidth]
{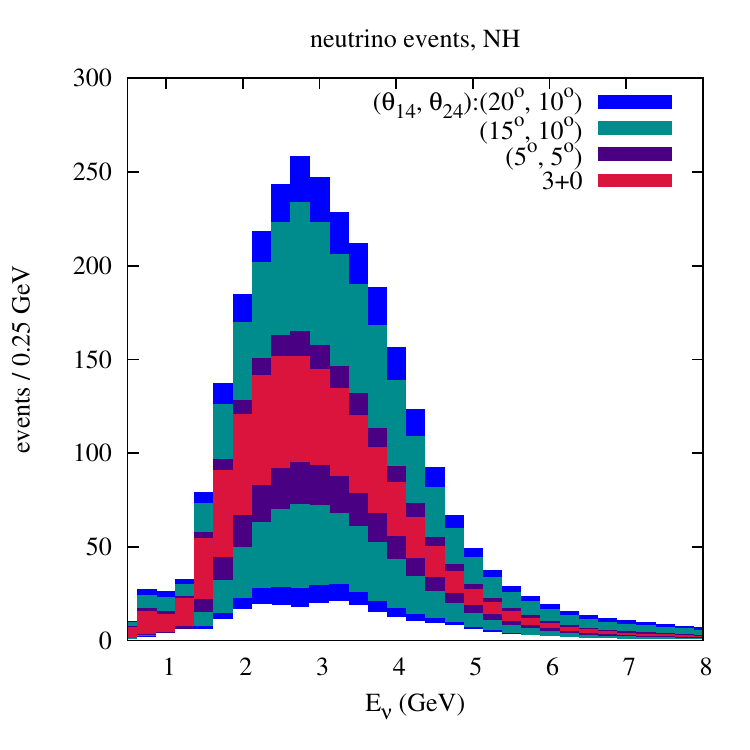}
\includegraphics[width=0.45\textwidth]
{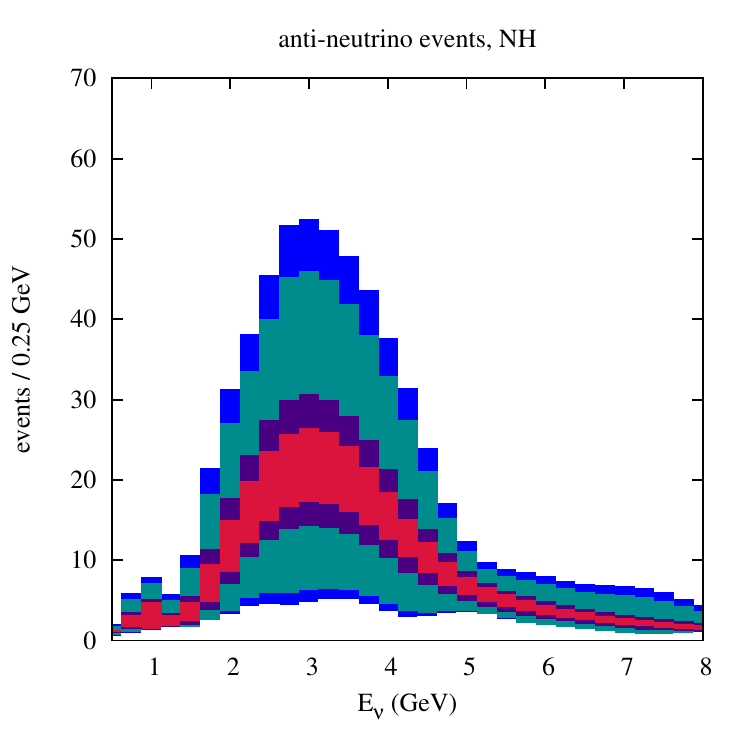}
\includegraphics[width=0.45\textwidth]
{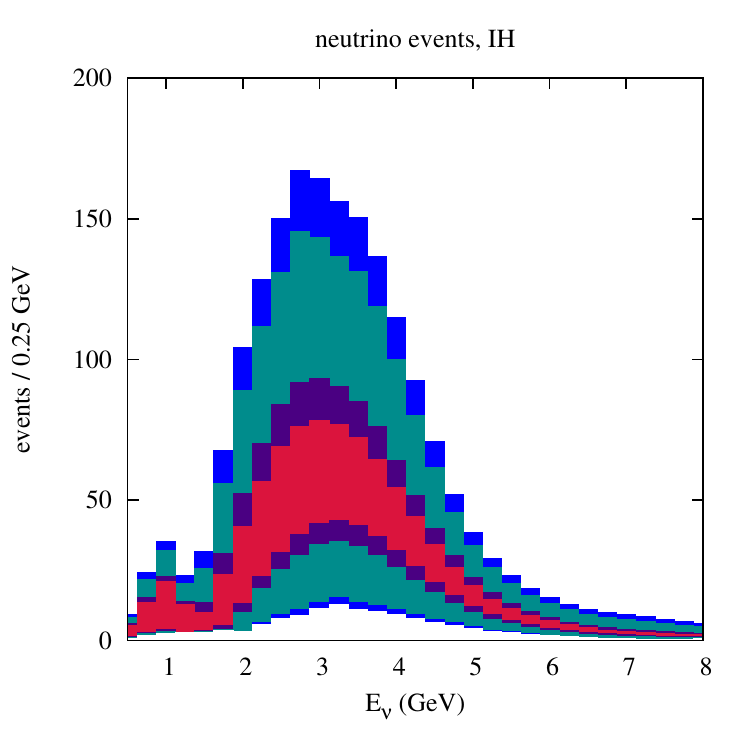}
\includegraphics[width=0.45\textwidth]
{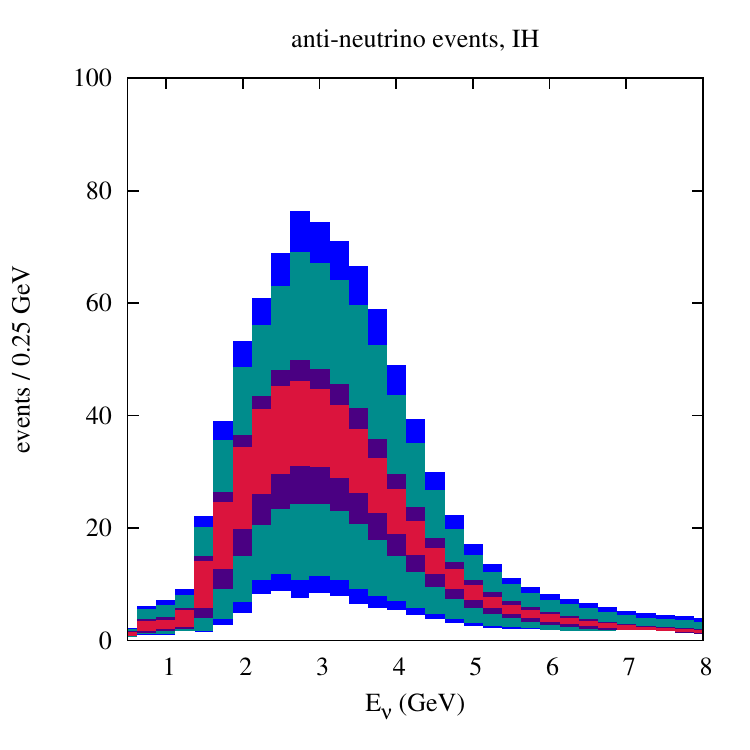}
\caption{\footnotesize{Neutrino and anti-neutrino event rates in DUNE plotted as a 
function of the reconstructed neutrino energy. 
The vertical spread for a given color for an energy bin shows the maximum 
and the minimum events rates possible. }}
\label{events_bands}
\end{figure}

Fig.\ \ref{events_bands} shows the spread of binned events as a function 
of the reconstructed neutrino energy for the 3+0 case and the 3+1 case.
For the 3+0 case, we varied only $\dcp$ in the range $[-180^\circ, 180^\circ]$
to obtain the events band shown in red. For 3+1, we chose three sets
of $\ta, \tb$ values - $(20^\circ, 10^\circ)$, $(15^\circ, 10^\circ)$
and $(5^\circ, 5^\circ)$. For all these three cases, we varied $\tc$ in the
range $[0, 30^\circ]$ and the phases $\da$, $\db$ and $\dc$ in the range
$[-180^\circ, 180^\circ]$ each. The resulting event-bands are shown in blue, green
and magenta, respectively. The left (right) panels  show the neutrino
(anti-neutrino) rates, while the top (bottom) panels are for the  NH (IH) scenario.

It can be seen that for all three sets of $\ta, \tb$, the 3+1 band can
potentially 
encompass the 3+0 band, leading to substantial degeneracy.  When the number
of events falls in the overlapping region between these two bands (which is the 
red region in Fig.\ \ref{events_bands}), there is 
considerable  ambiguity as to whether the events are produced by a certain value
of $\dcp$ in the 3+0 sector or by some combination of $\tc$, $\da$, $\db$
and $\dc$ in the 3+1 sector.

Fig.\ \ref{events_bands} also shows that the 3+1 band gets wider as the 
values of  $\ta, \tb$ and hence the effective mixing angle 
$\sin{2\theta_{\mu e}^{4\nu}}$ increase.  Indeed, for sufficiently large 3+1 mixing angles, the 3+1 band is substantially larger than its 3+0 counterpart. An observed surfeit
or  a dearth of events compared to those expected in the 3+0 case,
especially near the event maxima 
(around the region 2-4 GeV), could be a  pointer to the presence of sterile states. 

\section{Implications}
\label{sec:implic}

Prior to summarizing our conclusions, it is useful to  discuss certain
implications which arise from our results. We have noted above
that in the presence of even a single sterile neutrino, conclusions such as, a) Whether CP is conserved or violated, and b) if the latter, whether the violation is
ascribable to the active neutrinos  or the additional sterile neutrino,  or a
combination of the two, are all rendered significantly ambiguous.  An important  consequence  is thus 
the need for an improved synergistic linkage between the global LBL 
and SBL efforts \cite{Zennamo}, since it appears that results obtained in  the
former cannot be correctly interpreted without definitive conclusions  drawn
from the  latter.

  Our work also has ramifications for near detector (ND)  design and physics requirements in DUNE and other LBL experiments. Even if there is no sterile sector
and the standard three-family scenario is nature's choice, in order to fully exploit the CPV  capabilities of a far detector (FD), the ND must establish the expected number of events at the FD in the absence of oscillations with very high precision, in order to ensure systematic errors stay
well below statistical ones. The rates expected at the FD depend on fluxes and cross-sections measured, along with their energy dependence, to significantly high accuracy  at the ND for all four species of neutrinos, $\nu_e,
\bar{\nu}_{e}, \nu_{\mu}, \bar{\nu}_{\mu}$. In the $3+0$ scenario, these measurements, while very demanding, are assumed to be made under conditions where there are no oscillations between the source and the ND. This task is  rendered significantly more complex, however, in the presence of a sterile sector capable of altering the fluxes between the source and the ND over the planned distance of $\sim500$ m in DUNE. Historically, uncertainties in source fluxes and cross-sections  have always been a major limiting factor for neutrino experiments seeking  precision in oscillation  studies. The high intensity of the DUNE beam and the presence of a highly capable and precise ND, designed to overcome these limitations,  are thus rendered even more crucial than assumed earlier in order to deconvolute the added complexities arising due to a sterile sector, if it were to be present in nature.
Moreover, we note that current data allow the  mass-squared splitting in the sterile sector  to be sufficiently large such that oscillations due to it  can average out even at the ND distance of $\sim$500 m. Such an eventuality further underscores the need for a strong connection between the  SBL and LBL experimental programs, since these oscillations, even though not directly observed at the ND, would nonetheless affect its crucial service task of accurate flux determination.

\section{Summary and Conclusions}
\label{sec:summary}

To summarize, we have studied the effects of the additional mixing angles and CP
phases in the
case of a 3+1 sterile  sector on the determination and measurement  of CPV at
long baselines for  the DUNE experiment. From a probability analysis, we show
that the effects of the additional CP  phases can be large at its chosen
baseline of 1300 km. These effects, which arise from large interference terms
(between the 3+0 and 3+1 sectors) in the appearance probability, are accentuated
by the presence of matter, which additionally brings in contributions from
sterile-sector  mixings and phases which are dormant at short baselines. 
From event rate calculations, we show that the  presence
of a sterile sector manifests itself in measurably altered rates in energy bins
across the spectrum, without significant distortion in the shape. This
alteration in event rates increases, as expected, for larger values of the 
 mixing angles connecting the active and sterile sectors.

Importantly, the presence of a sterile sector obfuscates conclusive
determinations of CP violation or conservation at the far detector, and makes
uncertain the ability to ascribe any perceived CPV to a unique phase in the 3+0
sector.  Thus, the linkage between the presently planned long and
short baseline 
programs must be explored and strengthened. Until the presence of an $\sim \text{eV}^2$
sector is conclusively ruled out, our work emphasizes the need for a
complementary   SBL  sterile-search program and for a
highly capable and versatile  near detector  for DUNE, enabling it to reduce
systematics to low levels  so that it may  achieve its stated primary goals for
CPV detection.

\label{concl}

\section*{Acknowledgements}
\label{sec:ackg}

We thank Mary Bishai for help with DUNE fluxes and for patiently answering questions 
related to DUNE. RG and BK thank William Louis for very useful discussions. RG acknowledges useful discussions with Sandhya Choubey, Amol Dighe and S. Uma Sankar. MM thanks Animesh Chatterjee 
for very helpful discussions.  RG and BK are 
grateful to the Kavli Institute for Theoretical Physics for support and
hospitality. RG acknowledges support from the Fermilab Neutrino Division
and the DUNE Project at Fermilab while this work was in progress in the form of an Intensity Frontier
Fellowship in 2014 and a Guest Scientist position in 2015. RG, MM and SP also acknowledge support from the XII Plan 
Neutrino Project of the Indian DAE. BK thanks the Mainz Institute for Theoretical Physics 
for its hospitality and support while this paper was completed. This research was supported in part by the US National Science 
Foundation under Grant No. NSF PHY11-25915. Fermilab is operated by Fermi Research 
Alliance, LLC under Contract No. DE-AC02-07CH11359 with the US Department of Energy.

\appendix

\bibliographystyle{apsrev}
\bibliography{references.bib}

\end{document}